\documentclass[sigconf]{acmart}

\usepackage{xcolor}
\usepackage{booktabs}
\usepackage{array}
\newcommand*\rot{\rotatebox{90}}
\newcolumntype{L}[1]{>{\raggedright\arraybackslash}m{#1}}
\newcolumntype{C}[1]{>{\centering\arraybackslash}m{#1}}
\newcolumntype{R}[1]{>{\raggedleft\arraybackslash}m{#1}}

\AtBeginDocument{%
  \providecommand\BibTeX{{%
    \normalfont B\kern-0.5em{\scshape i\kern-0.25em b}\kern-0.8em\TeX}}}

\setcopyright{acmcopyright}
\copyrightyear{2020}
\acmYear{2020}
\acmDOI{10.1145/1122445.1122456}

\begin{document}

\title{Secure Federated Learning Approaches to Diagnosing COVID-19}

\author{Rittika Adhikari}
\affiliation{%
  \institution{University of Illinois at Urbana-Champaign}
  \city{Urbana}
  \state{Illinois}  % Optional, depending on the country
  \country{United States of America}
  }
\email{rittika2@illinois.edu}

\author{Christopher Settles}
\affiliation{%
  \institution{University of Illinois at Urbana-Champaign}
  \city{Urbana}
  \state{Illinois}  % Optional, depending on the country
  \country{United States of America}
  }
\email{csettl2@illinois.edu}

\renewcommand{\shortauthors}{Adhikari and Settles}

\begin{abstract}
The recent pandemic has delivered a desire for understanding COVID-19 diagnoses among patients in hospitals. Currently, hospitals have difficulty diagnosing this novel respiratory illness from a patient’s chest X-Ray to another, simply because it is difficult to compare COVID chest X-Ray between patients due to HIPAA compliance. In this paper, we aim to build a model to assist in the diagnosis of COVID-19 while being HIPAA compliant through federated learning, a distributed machine learning technique used to train an algorithm across multiple decentralized devices with local data samples without sharing them \cite{choudhury2020differential}. Our model extends on existing work in the chest X-Ray diagnostic model space; we analyze the best performing models in the CheXpert \cite{irvin2019chexpert} competition and build our own models that work effectively for our hospital data. Given that this model is being iterated upon in a federated setting, we consider the possibility of our model being updated using biased data and analyze its effect on the model's final performance. Additionally, in order to provide the respective hospitals with more insight into how the learned federated model makes its decision, and to ensure that the model is not learning some arbitrary unimportant features, we utilize Grad-CAM \cite{Selvaraju_2019}to visually emphasize which features on a patient leads to a positive COVID diagnosis.
\end{abstract}

\maketitle

\section{Introduction}
Since the novel COVID-19 virus first came to light in Wuhan Province, China, doctors and researchers alike have been searching for a solution to easily diagnose patients with COVID. Currently, COVID is typically detected using the reverse-transcription polymerase chain reaction (RT-PCR), as per the guidelines set by WHO \cite{sohrabi_alsafi_oneill_khan_kerwan_al-jabir_iosifidis_agha_2020}. However, this form of testing is both time consuming and risky for patients who are actually suffering from COVID-19, so this is typically prefaced with medical imaging, such as X-Rays and CT scans, to ensure that there are signs of COVID before continuing with the RT-PCR test.

Our paper proposes building a model through federated learning to automatically diagnose COVID from patient chest X-Rays efficiently and accurately. This model would aid as a useful supplement to doctors who are trying to quickly determine whether a patient needs to continue with the RT-PCR test from their chest X-Ray results. Initially, one might question why the model needs to be trained in a federated setting -- wouldn't it be enough to train separate models per hospital? However, our concern is that the data at a single hospital may be severely biased and would not be an accurate representation of all patient chest X-Rays across hospitals \cite{tommasi2015deeper}. Thus, we utilize federated learning techniques to ensure that the model will be able to generalize well to samples from the other hospitals. Additionally, since it is inherently possible for an attacker to learn personally identifiable information from a model, as can be seen in \cite{shokri2017membership, carlini2019secret, fredrikson_jha_ristenpart_2015}, we propose a methodology of secure aggregation which is robust to such privacy attacks.

The model architecture we utilize references existing work in the chest X-Ray diagnostic space; specifically we refer to ensembled model architectures \cite{pham2020interpreting, ye2020weakly}\footnote{https://github.com/jfhealthcare/Chexpert} that were the most performant in the CheXpert \cite{irvin2019chexpert} competition. As a preprocessing step, we cleanse our data of any additional artifacts on the images, which may get into the way of the classification. For the purposes of training the global model, we utilize a publicly available dataset known as COVIDGR \cite{tabik2020covidgr}\footnote{https://github.com/ari-dasci/OD-covidgr}. This dataset was pre-processed using a segmentation-based cropping methodology. While building our model, we also consider and analyze the possibility of the hospital data being heavily biased, leading to biased global model updates.

We also utilize Grad-CAM\cite{Selvaraju_2019} to create heatmaps using the intermittent global models to demonstrate and emphasize the features that were heavily weighted during classification. We hope that this layer of interpretability will allow doctors and other researchers to gain further insights into what might be relevant when looking for COVID-19 symptoms \& potential treatment options.

To summarize, our contributions in this paper are as follows: 
\begin{enumerate}
    \item Extend existing model architectures used in the chest X-Ray diagnosing problem space to diagnose the novel COVID-19.
    \item Create an end-to-end secure federated learning system which utilizes the aforementioned model architecture to diagnose COVID-19.
    \item Utilize a novel local post-process interpretability method for distributed machine learning, which will allow users to better visualize important features.
\end{enumerate}

\section{Related Work}
This paper builds off of existing work in the medical image diagnostic space and in the secure federated learning space. In this section, we discuss the most relevant papers in these respective ares.

\subsection{Medical Image Diagnostic Methods} 
Using machine learning to solve complex diagnostic problems is nothing new \cite{pesapane_codari_sardanelli_2018}. Machine learning has been in the forefront of a variety of diagnostic challenges. One challenge that is particularly relevant to this paper is CheXpert \cite{irvin2019chexpert}, a competition created by a group of Stanford researchers to encourage other researchers to build diagnostic solutions to detect a variety of respiratory illnesses. CheXpert \cite{irvin2019chexpert} is a large, public, labelled chest radiograph dataset, consisting of 224,316 chest radiographs of 65,240
patients labeled for the presence of 14 observations as positive, negative, or uncertain. Since the competition's inception in 2019, several researchers have come up with innovative solutions, some of which even outperform actual radiologists \cite{pham2020interpreting, biospace_2019, ye2020weakly}\footnote{https://github.com/jfhealthcare/Chexpert}. \cite{pham2020interpreting, ye2020weakly} both utilize a machine learning technique known as transfer learning to build off of existing model architectures. Transfer learning is a method which utilizes a model developed for a different, yet somewhat similar, task as the starting point for a separate task. 

\subsection{Secure Federated Learning Methods} 
Federated learning \cite{mcmahan2017communicationefficient} is a machine learning technique which was developed to train a model on a variety of sensitive data, while ensuring the privacy of the aforementioned data. Researchers at Google \cite{10.1145/3133956.3133982} first proposed a federated learning system to collaboratively learn a shared prediction model, using data from edge devices. This system is robust to client failures (a highly valued trait when dealing with clients that are mobile phones and frequently loose connection). Despite allowing for robustness, the algorithm they propose still maintains the privacy of individual 'contributors', which in many of Google's cases are cell phones. The proposed secure aggregation protocol utilizes shared masks between pairs of clients, which when completely aggregated, cancel eachother out, allowing the aggregator to only ever know the summed model, but never being able to derive any one particular client's model.
 
\subsection{Federated Learning with Non-IID Data} 
Independently identically distributed (IID) data is the property where all the data in a particular sample are taken from the same distribution, and are independent from one another. Mathematically, for $x^{(i)}, x^{(j)}\thinspace\forall\thinspace i \ne j$ within a sample $S$, each $x^{(i)} \sim D$ (identically distributed) and $P[x^{(i)}, x^{(j)}] = P[x^{(i)}]P[X^{(j)}]$ (independently distributed) \cite{noniidblog}. This particular assumption of data distribution allows for models across different clients to be simply aggregated, without incurring any noticeable performance loss.

However, though the conditions for IID data are ideal, we unfortunately cannot realistically guarantee that our data distribution will be IID. Non-IID data is the property where the data in a particular sample either violate the independence assumption or the identicalness assumption. Mathematically, for $x^{(i)}, x^{(j)}\thinspace\forall\thinspace i \ne j$ within a sample $S$, at least one of $x^{(i)} \not\sim D$ (not identically distributed) and $P[x^{(i)}, x^{(j)}] \ne P[x^{(i)}]P[X^{(j)}]$ (not independently distributed) holds \cite{noniidblog}. We will encounter a non-IID data distribution across clients the majority of the time. 

Li et. al. \cite{Li_2020} includes maintaining statistical homogeneity as one of the several challenges in federated learning. Zhao et. al. \cite{zhao2018federated} showed that the training of federated systems where clients contain non-IID data can cause a significant model quality loss. They propose a solution of adding some globally known data to each client to soothe the problem a bit. We investigate the severity of this problem in our work, and propose future methods to alleviate it.

\subsection{Federated Learning in Diagnostic Applications} The privacy-preserving and information sharing aspects of federated learning is a huge sell for medical applications. \cite{choudhury2020differential} Currently, hospitals cannot share patient-specific information across different hospitals, as this is in direct violation of HIPAA. Thus, it can be hard for one hospital to build a sufficiently large and diverse dataset, which is essential to training an adequate model. However, with federated learning, hospitals can build a model together which reaches 99\% of model quality achieved through centralized data, while remaining generalizable to other institutions outside of the federation. \cite{fedlearninghosp}

There are several papers which discuss utilizing federated learning for practical medical applications. Brisimi et. al. \cite{brisimi_chen_mela_olshevsky_paschalidis_shi_2018} utilized federated learning methods with healthcare applications, specifically regarding predicting hospitalizations for cardiac events. Xu et. al. \cite{fedlearningehr} discusses federated learning to combine information from electronic health records, while maintaining patient privacy.

\section{Methods}
\begin{figure*}
  \includegraphics[width=\textwidth,height=3.5cm]{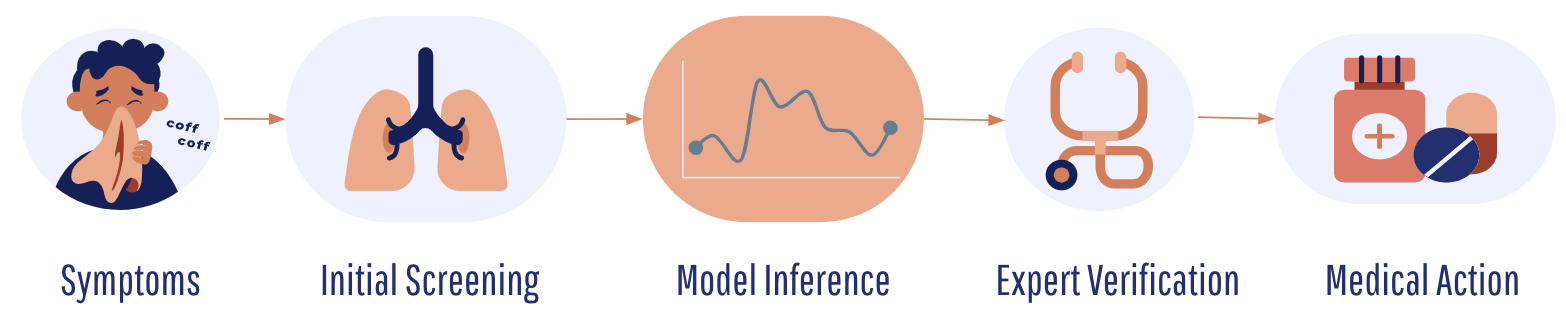}
  \caption{This figure describes the ideal end-user pipeline of our final secure federated learning system. First, the patient feels \textit{symptoms} of the illness and sets up an appointment at the doctor's office. Next, the patient goes to the radiologist and undergoes an \textit{initial screening}. We then utilize \textit{model inference} to diagnose the patient's X-Ray. Next, an \textit{expert verifies} the correctness of the diagnosis by analyzing an interpretable heatmap of the patient's X-Ray provided by the model. Finally, the expert prescribes the appropriate \textit{medical action} to begin to treat the patient's illness.}
  \label{fig:infra_overview}
\end{figure*}

We propose a methodology to build an end-to-end secure federated learning pipeline which can be utilized to diagnose COVID-19 and other respiratory illnesses using just a patient's X-Ray (see Figure \ref{fig:infra_overview}). There are two major components to our system - the machine learning model architecture used for diagnosis and the secure federated learning infrastructure built to support the aforementioned architecture.
\subsection{Model Architecture}
To build an effective model to diagnose COVID-19 from patient X-Rays, we referenced models \cite{pham2020interpreting, biospace_2019, ye2020weakly} that were the most performant on CheXpert \cite{irvin2019chexpert}, a large, public, labelled chest radiograph dataset. Several of these models utilized a technique called transfer learning, where they update the weights of an existing model trained to perform a similar task. Some of them also utilized a technique known as finetuning, where they freeze the first $n$ layers of the existing model, and then only update the non-frozen layers of the model. In designing our final model architecture, we explored both transfer learning and finetuning on DenseNet-121, DenseNet-169 \cite{huang2018densely}, and ResNet-18 \cite{he2015deep}. From preliminary experiments on COVIDGR, DenseNet-121 has the highest AUC when transfer learned or finetuned. Thus, we decided to base our model architecture off of DenseNet-121; when finetuning, we froze all layers except the last.

\subsection{Secure Federated Learning Infrastructure}
When building our federated infrastructure, we design for an honest-but-curious security setting \cite{6375935} where the following is maintained: 
\begin{enumerate}
    \item Each client (hospital) only has access to its own data, and it cannot observe other client's data. 
    \item Each client and the coordinator device exchange information such as to only share the global model after each local epoch has run. The coordinator or other hospitals can view the global model and attempt to glean results from it, but can never view another individual hospital's local model that is sent back to the coordinator in the encrypted form.
\end{enumerate}

We note that this setting of sharing the global model is an implementation of secure multi-party computation (first introduced by \cite{10.1007/978-3-540-88313-5_13}, and later applied to machine learning models in \cite{kumar2020cryptflow}, among others) where clients participating in the group collaborate to learn a secret value (the average of all their local models) while not having to explicitly share their local models. 

\subsubsection{Secure Aggregation}
Our secure aggregation protocol can be described as follows:
\begin{enumerate}
    \item Each pair of clients (hospitals) $i,j$ create a secret key $k_{i, j}$.
    \item When it comes time for model encryption, we imagine the machine learning model as an array of bytes. Each client (wlog, suppose client $i$) computes $K_{i,j} = PRNG(k_{i,j}, X) \forall j \ne i$, where $PRNG$ is a function which returns the next $X$ pseudo-random generated bytes from the key $k_{i,j}$, where $X$ is the 'length' of the ML model array. \footnote{In practice, we actually do this same process, but instead of 'imagining' doing it on an array, we do this process on every layer (array, or rather, torch tensor to be more exact) of the PyTorch model.}
    \item Client $i$ wishes to send $M^{\prime}_{i}$ and as such computes $$M^{\prime}_{i} + \sum_{j=i+1}^{n} K_{i,j} - \sum_{j=1}^{i-1} K_{i,j}$$ and sends this to the coordinator "aggregator" machine. 
    \item The coordinator machine sums together the "long byte array" models received from each hospital, then divides each layer of the model by the number of machines that were summed. This effectively averages the models together, cancelling out the individual masks that each hospital had applied, leaving only the average of the local hospital's models left to observe. 
\end{enumerate}

\begin{figure}
  \includegraphics[width=9cm, height=5.5cm]{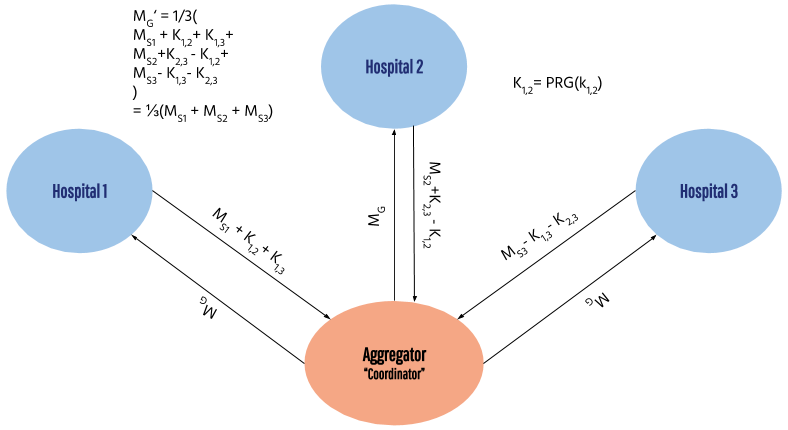}
  \caption[]{An illustration of the secure aggregation protocol. The coordinator machine sends the global model to each client, and each client responds with an encrypted version of the updated model. From the coordinator's perspective, the coordinator cannot decipher any individual model that is sent from a hospital, and can only learn the average of all the models by summing the models together to cancel out the encryption masks. }
  \label{fig:secure_agg_arch}
\end{figure}

Figure~\ref{fig:secure_agg_arch} provides an illustration of this protocol. 

\paragraph{Attacks against ML models}
Secure aggregation is an important factor in the security of our system design. As can be observed in \cite{geiping2020inverting}, federated learning systems that do not implement secure aggregation, and thus send back gradients from their clients to the server in the clear, have been shown to be vulnerable to the model inversion attack (first introduced in \cite{fredrikson_jha_ristenpart_2015}, and later improved upon in \cite{carlini2019secret}). Other attacks on machine learning systems include membership inference \cite{shokri2017membership}. These attacks seek to infer understanding of the underlying data that the local models were trained on, which to us presents a large concern since the data that our FL system is being built to work with is meant to be protected heavily. 

Note however that secure aggregation will only protect an adversary from viewing individual models sent back from the hospital - there is however an on-going problem of defending against attacks that can happen on the global model that the coordinator and every other machine in the system observe. Differential Privacy \cite{choudhury2020differential} is often used as a counter-measure in these scenarios to mask any original data that could have been learned from the model. Although this is a security concern on our part, we note that adding differential privacy to a FL system increases security, but also decreases utility of the models, as the models that are recovered from clients have random noise that disrupts the performance of the global aggregated model. We accept this security vulnerability, as we note that current standards for handling healthcare data require even less than what we have implemented already with secure aggregation. We also conjecture that training our models on chest X-ray images would reveal little to no details about a patient's personally identifiable information. 

\section{Experimental Setup}
In order to extensively evaluate our secure federated system to diagnose COVID-19, we ran experiments focusing on three different aspects. First, we assessed the reasonableness of the model using \textit{interpretability}. Next, we performed a \textit{sanity check} to ensure that our secure federated system was not incurring a significant performance loss compared to a mock insecure system. Finally, we measured the effect of the data distribution across clients to gauge the importance of an IID data distribution on our final global model.

\subsection{Interpretability Check}
To ensure that our model is making decisions in a logical manner, we utilized Grad-CAM \cite{Selvaraju_2019}, an interpretability technique used to produce a "visual explanation" for decisions made by convolutional neural networks. Selvaraju et. al.'s approach obtains a map which highlights only the features with a positive influence on the actual predicted class by utilizing the gradients of the activations with respect to the last convolutional layer, global average pooling, and the ReLU activation function. In this experiment, we utilize Grad-CAM to generate interpretable heatmaps on true positive, false positive, true negative, and false negative images, using a finetuned DenseNet-121 trained on 2 different clients over 20 epochs in an IID data distribution from COVIDGR \cite{tabik2020covidgr}. We then analyze these heatmaps to ensure that the highlighted features are all in the lung region, and aren't referring to some arbitrary region. In the future, we plan on discussing these heatmaps with medical experts to ensure that the highlighted features within the lungs are actually relevant to the classification of COVID vs. non-COVID.

\subsection{Sanity Check}
This experiment is conducted to ensure that our secure aggregation protocol does not hinder the ultimate performance of the model. We assess this by building a mock federated setting and comparing it to our true federated model setting through a comparison of ROC/AUC curves and confusion matrices. The mock federated setting setup simulates running $n$ different clients by training $n$ different models serially \footnote{We attempted training the models in parallel on our machine, however we observed that CUDA memory would often exhaust when more than 1 model was being trained at once.} on $n$ non-intersecting splits of data, and then aggregating them every $k$ epochs. In both the mock and true federated settings, we trained a finetuned DenseNet-121 using SGD with a learning rate of 0.01, 20 global epochs, and 2 clients with IID data collected from COVIDGR \cite{tabik2020covidgr}.

\subsection{Effect of Data Distribution}
As mentioned earlier, non-IID data has been proven to have a negative effect on the ultimate global model performance \cite{Li_2020, zhao2018federated}. We decided to conduct a few experiments to truly measure the effect of non-IID data on our ultimate global model performance. We trained three different models in order to gauge the effect of client data distribution on the final global model quality. To build these models, we transfer learned DenseNet-121 across 5 different clients using SGD with a learning rate of 0.01 and 20 global epochs. The only difference in the training settings was the type of data distribution across clients. Due to the small size of COVIDGR \cite{tabik2020covidgr} and the unfeasibleness\footnote{We consider these two datasets to be an unrealistic representation of true hospital data, as these images have been extracted from papers and are of very low quality. Thus, it is usually non-trivial to determine the decision boundary between COVID and non-COVID images due to the large gap in quality.} of the DeepCOVID \cite{minaee2020deepcovid} and COVID Chest X-Ray Dataset \cite{cohen2020covidProspective}, we instead predicted Cardiomegaly, an alternative illness, from the CheXpert \cite{irvin2019chexpert} dataset. We first reformatted CheXpert into a binary dataset, with Cardiomegaly as the positive class and all other images as a negative class.
The three data distributions were constructed from CheXpert as follows:
\begin{enumerate}
    \item \textbf{IID-Data} Evenly \& randomly split reformatted dataset into 5 train-test sets.
    \item \textbf{Non-IID Data} Randomly split reformatted dataset into 5 train-test sets by age range. The age ranges for the splits were 0-30, 31-44, 45-58, 59-72, 73-86, 87-90 for clients 1-5 respectively. This ensures that the data distribution across clients are not identical.
    \item \textbf{IID-Data w/ Different Train \& Test Distributions} Split reformatted dataset by age range such that all training samples are from age range 60-90 and all testing samples are from ages 0-59. Then, evenly split the train and test data into 5 non-intersecting sets.
\end{enumerate}

We evaluated the final global models on the full test dataset and generated ROC/AUC curves and confusion matrices for comparisons.

\section{Results and Discussion}
\begin{table*}\sffamily
    \begin{tabular}{ L{0.5cm} C{6.7cm}  C{6.7cm} }
        \toprule
         & COVID-19 & Non-COVID-19 \\ 
        \midrule
        \rot{COVID-19} & \includegraphics[width=6cm]{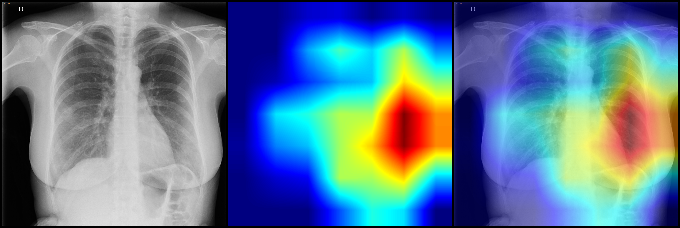} & \includegraphics[width=6cm]{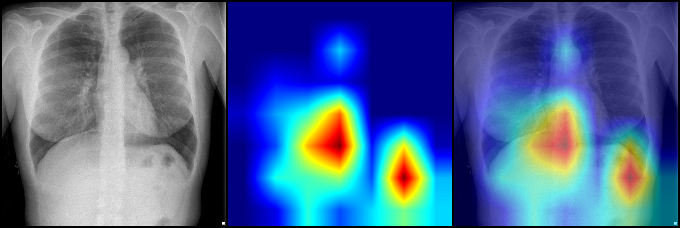}  \\
        \midrule
        \rot{Non-COVID-19} & \includegraphics[width=6cm]{images/fn_covid_heatmap.png} & \includegraphics[width=6cm]{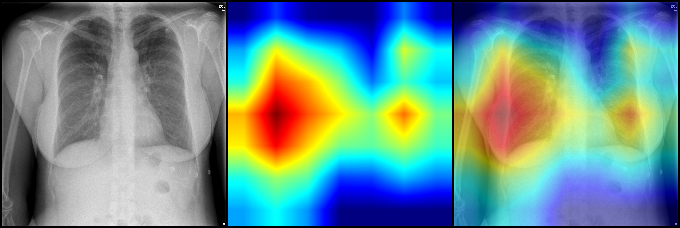}  \\ 
        \bottomrule 
    \end{tabular}
    \caption{Grad-CAM \cite{Selvaraju_2019} generated heatmaps on final global model from the \textit{Interpretability Check} experiment. \textit{Top left: }True positive. \textit{Top right: }False negative. \textit{Bottom left: }False positive. \textit{Bottom right: }True negative.}
    \label{tab:interp_check}
\end{table*} 
In this section, we discuss the results of the interpretability check, sanity check, and data distribution experiments. 

\subsection{Interpretability Check}
Our results from using Grad-CAM indicate that our model appears to be utilizing components from the lung portion of the X-ray image, providing us with confidence that our model is learning the appropriate features needed to identify COVID-19. Though we do not have a professional opinion on the quality of the heatmaps produced by Grad-CAM \cite{Selvaraju_2019}, we are optimistic of our model's feature learning. 

\subsection{Sanity Check}
We observe that the performance of our Mock setting and our true setting match pretty closely (with AUC scores of 0.819 and 0.825 of the mock and true settings respectively). This gives us confidence that our implementation of our federated learning system is correct, and as a result, we are able to use it to train models in a distributed setting, speeding up the runtime needed to perform training tasks for our next experiments. 

\begin{table*} \sffamily
    \begin{tabular}[t]{ L{0.5cm} C{3.7cm}  C{3.7cm} }
        \toprule
         & ROC Curve & Confusion Matrix \\ 
        \midrule
        \rot{Mock Setting} & \includegraphics[width=3.5cm]{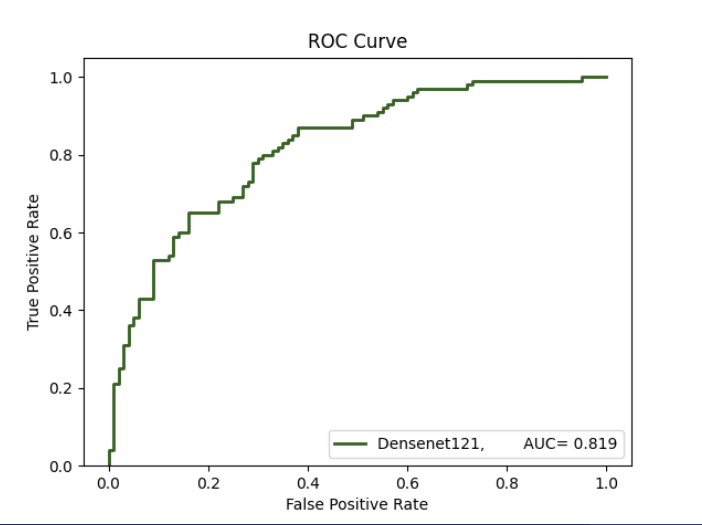} & \includegraphics[width=3.5cm]{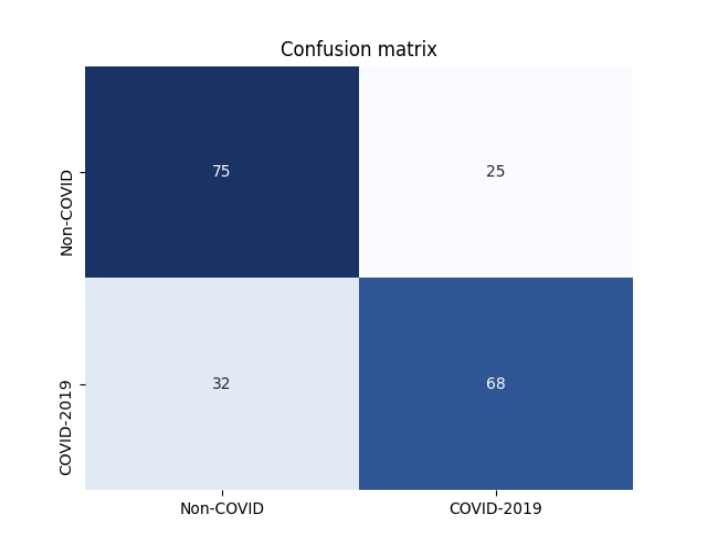}  \\ 
        \midrule
        \rot{True Setting} & \includegraphics[width=3.5cm]{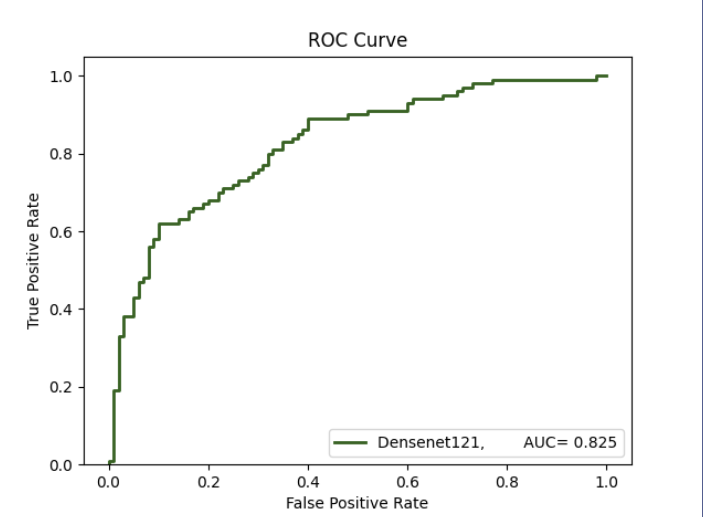} & \includegraphics[width=3.5cm]{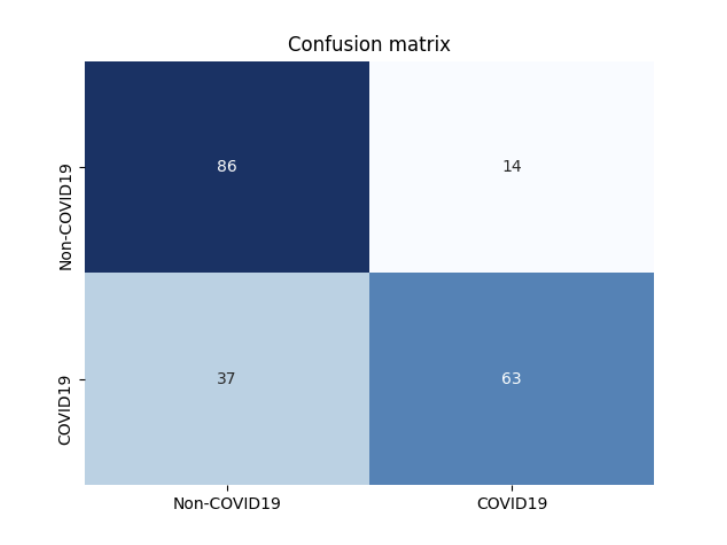}  \\ 
        \bottomrule 
        \label{tab:sanity_check}
    \end{tabular}
    \caption{Comparison of ROC/AUC curves and confusion matrices from models trained through a mock federated setting versus the true secure federated setting.}
\end{table*}

\subsection{Effect of Data Distribution}
We assessed the effect of the performance of training global models on three different types of data distributions - IID data, non-IID data, and IID data with differing train \& test distributions. 

\subsubsection{IID Data}
The global model trained on IID data acts as our "control". IID data, as mentioned previously, is our ideal scenario of data distribution across clients. Here, we can aggregate the models trained on each client without worrying about one of them heavily biasing the final global model. As can be seen in Table \ref{tab:data_dist}, we obtain a relatively high AUC of 0.826, with a high concentration of true positives and true negatives. 

\subsubsection{Non-IID Data}
The global model trained on non-IID data is our "realistic" situation. It is unlikely that we will attain IID data across clients, since the demographics of patients will be different from hospital to hospital. For example, there might be skews in the racial, age, or gender demographics. Additionally, there might be a skew in socioeconomic backgrounds, which does tend to play a role in healthcare situations. As demonstrated in Table \ref{tab:data_dist}, though the AUC remains similar to that of the IID data model at 0.827, there is a significantly higher concentration of false negatives, which could be fatal. 

\subsubsection{IID Data with Different Train \& Test Distributions}
The global model trained on IID data with different train and test distributions is an experimental situation. It is within plausibility for this situation to occur, so we decided to explore this route. In this situation, the clients all have IID train data and test data from a different IID distribution. In Table \ref{tab:data_dist}, it is evident that this has had a severely negative effect, as the AUC has dropped to 0.804. Additionally, the number of false positives has increased compared to that of the model trained on regular IID data. This is not as disastrous as a larger number of false negatives; however, depending on the treatments required, this could lead to unnecessary intrusive and painful treatments.

\begin{tabular}[t]{ L{0.5cm} C{3.7cm}  C{3.7cm} }
    \toprule
    & ROC Curve & Confusion Matrix \\ 
    \midrule
    \rot{IID Data} & \includegraphics[width=3.5cm]{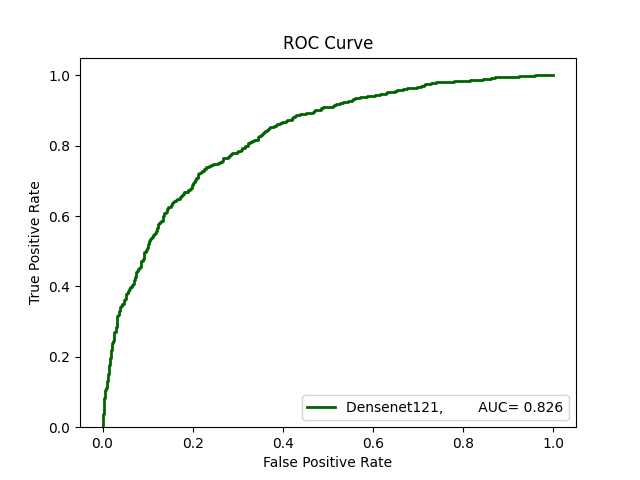} & \includegraphics[width=3.5cm]{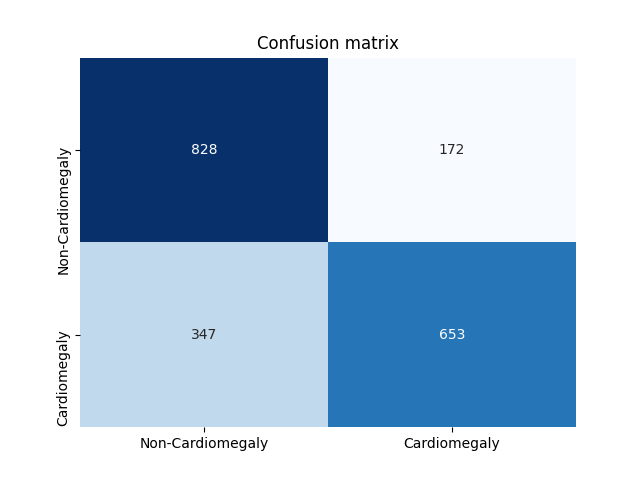}  \\
    \midrule
    \rot{Non-IID Data} & \includegraphics[width=3.5cm]{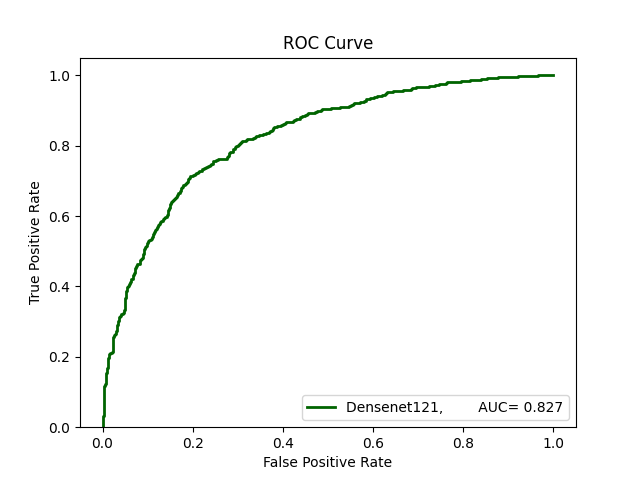} & \includegraphics[width=3.5cm]{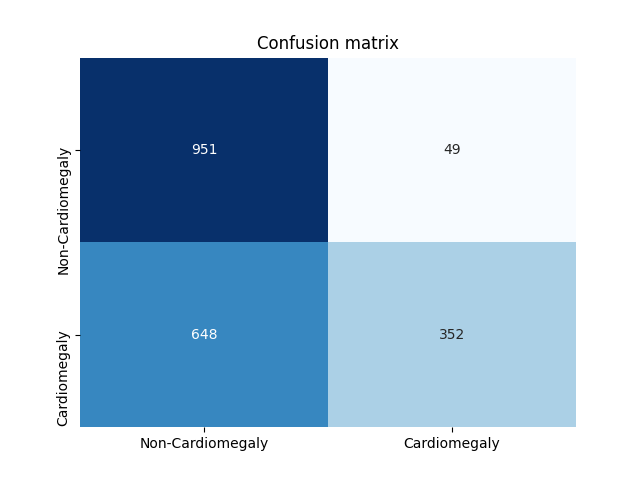}  \\
    \midrule
    \rot{IID Data, Diff Trn/Tst} & \includegraphics[width=3.5cm]{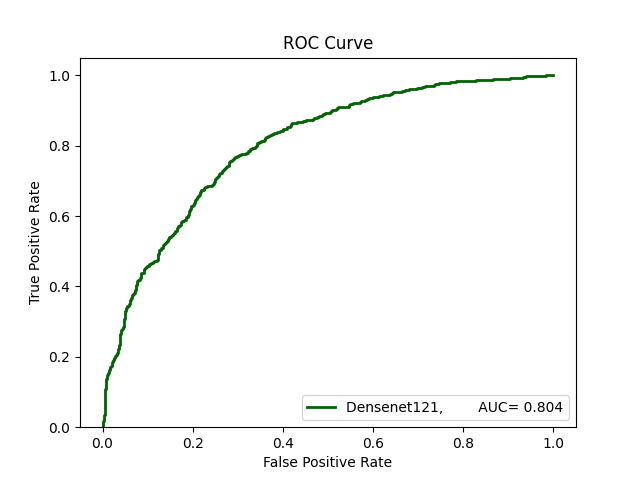} & \includegraphics[width=3.5cm]{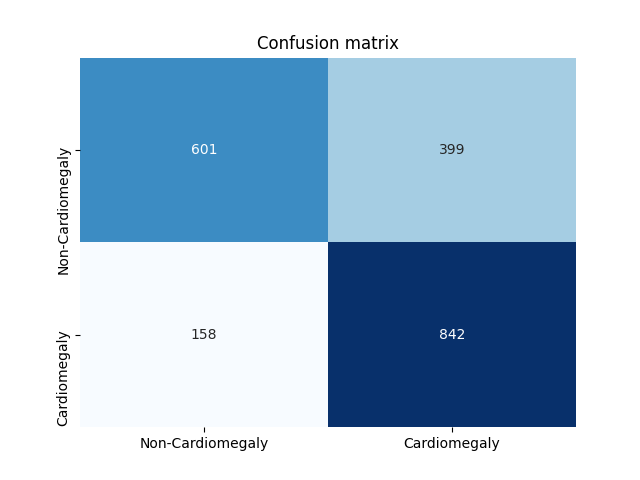}  \\
    \bottomrule 
    \label{tab:data_dist}
\end{tabular}
\captionof{table}{Comparison of ROC/AUC curves and confusion matrices from models trained on IID data, Non-IID data, and IID data with different train \& test distributions.}

\section{Conclusions}
In this work, we present a federated learning system that classifies X-Ray images into COVID-19 or non-COVID-19, which preserves privacy of individual model updates by utilizing secure aggregation techniques. Our methodology extends prior work in secure aggregation in federated learning and references model architectures that were the most performant in the CheXpert competition. We utilize a local post-process interpretability method, Grad-CAM, to generate interpretable heatmaps which allow medical experts to better visualize important features and hopefully glean relevant information. We evaluate our method through a series of experiments to ensure that our secure federated system does not compromise our final model performance compared to a non-secure centralized version. We also conduct experiments to measure the effect of the data distribution on the efficacy of our federated system. 

From our preliminary experiments on the effects of non-IID data on our federated system, we would like to explore utilizing an algorithm such as FedProx \cite{li2020federated} or Scaffold \cite{karimireddy2020scaffold} which utilizes forced convergence to penalize deviation from the global model to ensure that the final model is able to generalize well and remains performant. Additionally, we would like to add a pre-processing step to align the X-Rays in a uniform manner, similar to how Tabik et. al. \cite{tabik2020covidgr} utilized segmentation-based cropping. We would consider segmentation techniques as detailed in \cite{badrinarayanan2016segnet, ronneberger2015unet}. We hope to investigate a technique known as ensembling to combine multiple different learners to create a better predictor. \cite{pham2020interpreting, ye2020weakly} both use simple averaging to combine the learners; we propose to go one step further and use weighted averaging \cite{davila_mobin_1992, 9121222}. Eventually, we want to improve our model to be able to distinguish the severity of the illness, given appropriate data. We also plan on exploring potential privacy attacks \cite{9109557} on our federated learning system, and how we can make it more robust to such attacks. We also look forward to receiving real hospital data from OSF Healthcare and collaborating with them to make their systems more performant.

\begin{acks}
We would like to thank Professor Sanmi Koyejo, Professor Dakshita Khurana, Derek Xia, Vikram Duvvur, Sam Pal, Raj Velicheti, Nishant Kumar, and Jacky Yibo Zhang for all of their help and guidance, without whom this work would not be possible. We would also like to thank Professor Gang Wang for all of his valuable feedback throughout this research project. We also thank the C3.AI Digitial Transformation Institute (DTI), for providing us funding for our research, and OSF Hospital for their current and future collaboration with our group. 
\end{acks}

\bibliographystyle{ACM-Reference-Format}
\bibliography{federated}

\appendix 
\section{Federated Learning - Engineering Design}
From an engineering point of view, we provide some more details on our implementation. 
    \paragraph{Using docker containers} Each hospital downloads a docker container we provide for them, and they run the container. This detail is important as it emphasizes the simplicity that we want hospitals to have when on-boarding our system. Using docker containers allows hospitals to not have to worry about environment setup issues, and instead just worry about about running the docker container that we provide to them on a public-internet-facing machine. 
    
    \paragraph{gRPC server API} The container runs a gRPC server that exposes three endpoints:
    \begin{itemize}
        \item An \texttt{Initialize} RPC which takes a list of other machines participating in the FL system, and initiates the key exchange protocol by connecting to every other machine (by calling another endpoint called \texttt{GenerateSharedKey} that has a larger IP address, and asking the machine to generate a shared key between the pair for the rest of the training rounds.
        \item The previously mentioned \texttt{GenerateSharedKey} method generates 1024 bits of randomness from \texttt{/dev/random}, and uses this as the shared secret key for all future psuedo-randomness exchanged between machines. Note that this exchange of the shared key happens over TLS, which is implemented via gRPC secure channels. 
        \item A \texttt{ComputeUpdatedModel} RPC which takes the global Pytorch model, dumps the data to a stream of bytes using python's pickle library, and sends the pickled model in the gRPC request to each of the clients. 
    \end{itemize}
    \paragraph{Converting floating point to fixed point}
        We note that, because PyTorch layers are effectively tensors of floating point numbers, sending 'encrypted' versions of them which can be decrypted by summing the values together will not work in the case of floating point numbers because the floating point representation of the binary (bit-for-bit) addition of two floating point numbers is not equal to the sum of the two floating point numbers. Thus, we solve this problem by converting all the weights on the hospitals to fixed-point (integer precision). This is done by effectively taking the floating point representation of the model parameters and multiplying them by a constant scalar $2^{24}$, and then converting the numbers to integers. This technique was shown effective in \cite{kumar2020cryptflow}, and effectively preserves a large part of the precision of the floating point number by first shifting the number up before casting it to an integer. Then, on the coordinator side, after we sum all the models together, we then divide every model parameter by the same scalar $2^{24}$. \footnote{$2^{24}$ was found to be an effective shift amount empirically, however other values could also be used.}
    \paragraph{Global and local epochs}
        After the coordinator sends the global model to each hospital, each hospital runs \textit{n} local epochs, and sends the updated and encrypted model to the coordinator.
        On receipt of the hospital weights, the coordinator securely averages them to build one new global model, and then repeats step 2 for \textit{e} global epochs.
        Once \textit{e} global epochs have completed, global model is saved on the coordinator and can be used in similar ways for inference among the federated learning clients. 

\begin{figure}
  \includegraphics[width=9cm]{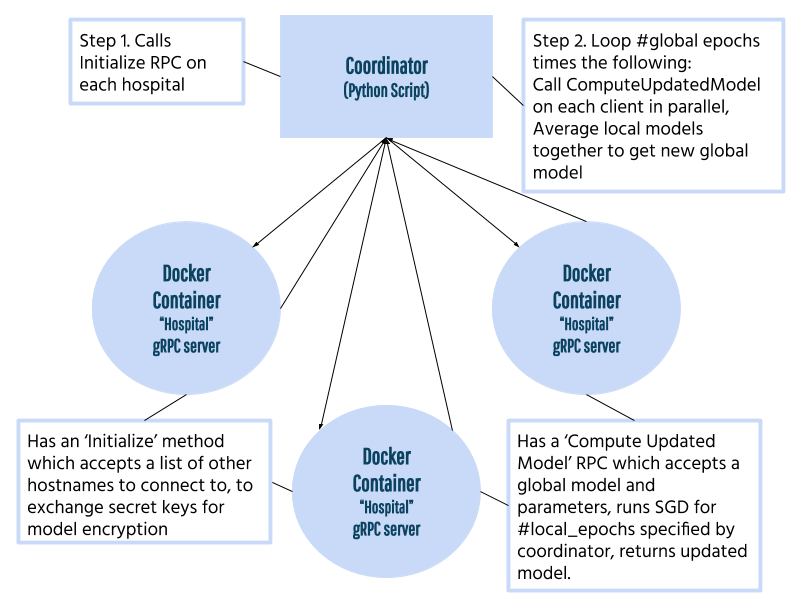}
  \captionof{figure}{} An illustration of the implementation of the FL system.
  \label{fig:fl_distsys_arch}
\end{figure}

See Figure \ref{fig:fl_distsys_arch} for a more detailed overview of our federated system.

\end{document}